# An HST/WFPC Survey of Bright Young Clusters in M31. II. Photometry of Less Luminous Clusters in the Fields

P. W. Hodge[1], O. K. Krienke[2], M. Bellazzini[3], S. Perina[3], P. Barmby[4], J. G. Cohen[5], T. H. Puzia[6], and J. Strader[7]

[1] Department of Astronomy, University of Washington, Seattle, WA 98195-1580, USA
[2] Seattle Pacific University, Seattle, WA 98119, USA
[3] INAF-Osservatorio di Bologna, via Ranzani 1, 40127 Bologna, Italy
[4] Department of Physics and Astronomy, University of Western Ontario, London, ON N6A 3K7, Canada
[5] Palomar Observatory, Mail Stop 105-24, California Institute of Technology, Pasadena, CA 91125, USA
[6] Herzberg Institute of Astrophysics, 5071 West Saanich Rd., Victoria, BC, V9E 2E7, Canada
[7] Harvard-Smithsonian Center for Astrophysics, Cambridge, MA, 01238, USA

**ABSTRACT.** We report on the properties of 89 low mass star clusters located in the vicinity of luminous young clusters ("blue globulars") in the disk of M31. 82 of the clusters are newly detected. We have determined their integrated magnitudes and colors, based on a series of *Hubble Space Telescope* Wide Field/Planetary Camera 2 exposures in blue and red (HST filters F450W and F814W). The integrated apparent magnitudes range from F450W = 17.5 to 22.5, and the colors indicate a wide range of ages. Stellar color-magnitude diagrams for all clusters were obtained and those with bright enough stars were fit to theoretical isochrones to provide age estimates. The ages range from 12 Myr to >500 Myr. Reddenings, which average E(F450 – F814) = 0.59 with a dispersion of 0.21 magnitudes, were derived from the main sequence fitting for those clusters. Comparison of these ages and integrated colors with single population theoretical models with solar abundances suggests a color offset of 0.085 magnitudes at the ages tested. Estimated ages for the remaining clusters are based on their measured colors. The age-frequency diagram shows a steep decline of number with age, with a large decrease in number per age interval between the youngest and the oldest clusters detected.

## 1. INRODUCTION

This paper reports on the study of open (disk) star clusters in M31 (NGC224) detected on images from the *Hubble Space Telescope*, obtained as part of a program designed to determine the nature of 19 luminous star clusters that were originally classified as globular clusters, but which have blue measured colors. The first paper of a series that report on the results of that program concerns the highly luminous young cluster vdB0 (Perina et al. 2009). Here we present a survey of less-luminous ("open") clusters in M31, similar to those of Krienke and Hodge (2007, hereafter KHI), who reported results from archival images obtained with the Wide Field/Planetary Camera 2 (WFPC2), and Krienke and Hodge (2008, hereafter KHII), who reported results from archival images from the Advanced Camera for Surveys (ACS).

"Open" or "disk clusters" in M31 have been recognized since Hubble's pioneering work. He identified the cluster subsequently known as vdB0 as an open cluster, as shown in the frontispiece of his book, "The Realm of the Nebulae" (Hubble 1936). Most subsequent studies of such clusters have dealt with the more luminous examples, especially those mistaken for globulars; see an excellent history of the subject of M31's luminous blue clusters in Caldwell et al. (2009).

As in Paper I, we adopt a distance modulus for M31 of $(m - M)_0 = 24.47$ +/- 0.07.

## 2. OBSERVATIONS

### 1 The Images

The observations, obtained with the Wide Field Planetary Camera 2 (WFPC2) of the *Hubble Space Telescope (HST),* were described in detail in Perina et al. (2009). The images were obtained with blue (HST F450W) and red (HST F814W) filters, approximately in the traditional B and I bands. Exposures were relatively short (2x400 seconds per filter). The scale of the WF fields is 0.099 arcsec/px and for the PC fields it is 0.045 arcsec/px. While the main program dealt with the bright globular-like clusters on the PC images, we searched both the PC and the WF images, identifying star clusters, measuring their integrated properties and carrying out stellar photometry of their member stars. Figure 1, in a color version produced by one of us (TP), reproduces a sample WF field showing several open clusters. The total area covered by the survey is 48.1 arc minutes$^2$.

### 2 Cluster Identification

The clusters included in the survey range from large, very luminous clusters to small objects that are barely resolved in our rather short exposures. The brightest disk clusters in this sample have absolute magnitudes of $M(F450)_0 = -8$, while we were able to identify a few clusters as faint as $M(F450)_0 = -2.5$. Thus our brightest clusters are equivalent to the mean absolute magnitudes of M31's globular clusters (though bluer and less massive), while our faintest are fainter than the faint limit of most cluster catalogs for nearby galaxies.

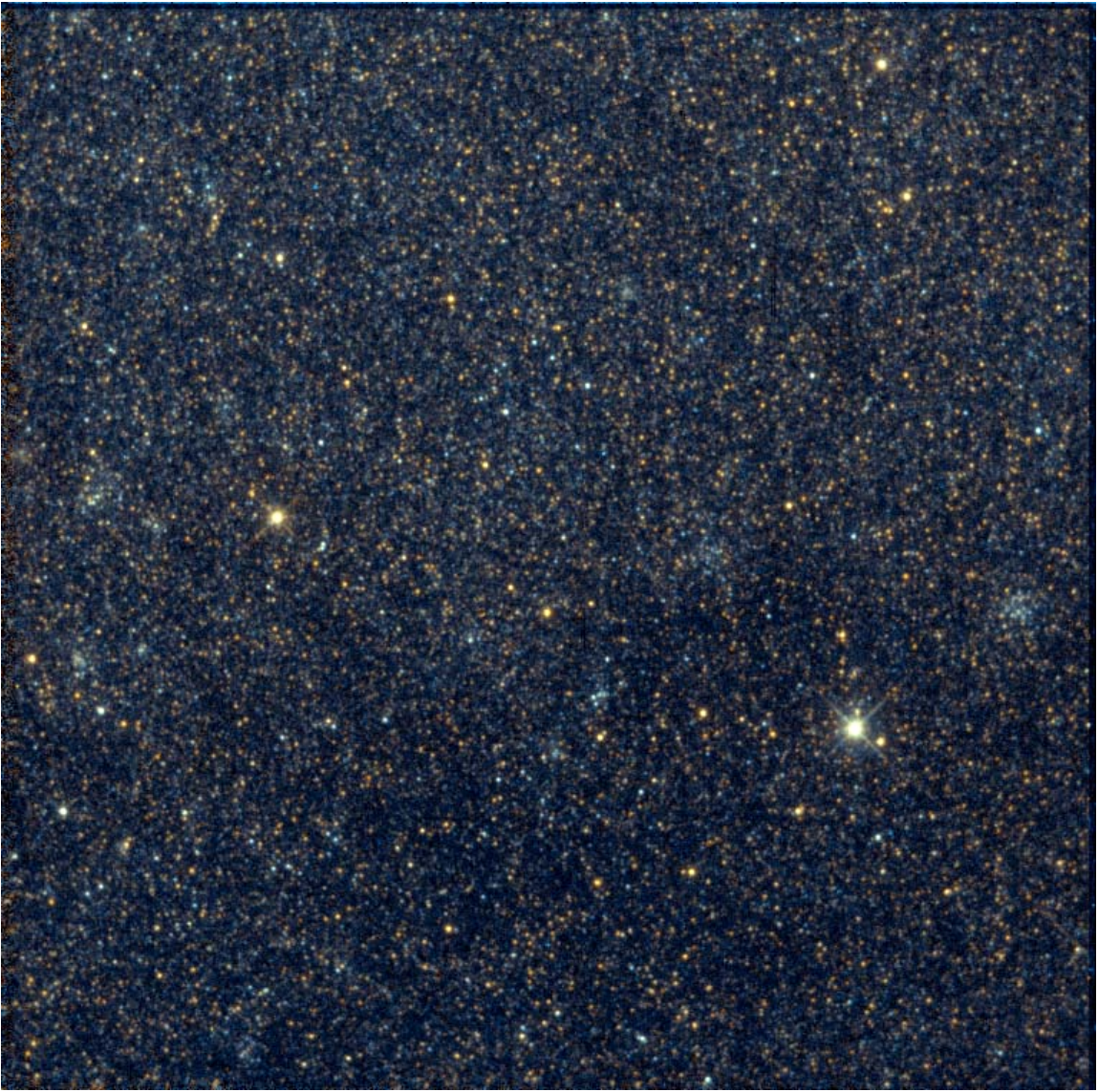

Figure 1. A sample WF image, containing several recognizable star clusters. This figure demonstrates how clusters are distinguished by their resolution, high stellar density and blue color, compared to the background of the M31 disk stars.

The disk of M31 presents a dense star field, in which low-density star clusters are difficult to detect even with special statistical techniques. For that reason we chose to select only conspicuous objects for which there would be little or no question of their being physical clusters (see examples in Figure 2). Our cluster identification criteria included:
1. a conspicuous spatial concentration
2. a centrally-peaked radial distribution
3. detectability in both colors
4. recognition of more than four well-resolved stars above an unresolved background

5. a normal luminosity distribution (number increasing with magnitude)
6. a color-magnitude diagram that shows a distribution different from that of the background

Two of the authors (PH and OKK) searched the frames independently in both colors, varying brightness and contrast. We categorized objects as definitely clusters or as candidates, and for borderline cases, we met, discussed images, and reached agreement. As a final test, we asked each other whether we could defend an object against being classed as an asterism, background galaxy or other type of non-cluster. Figure 2 provides F450 images of 12 of the clusters.

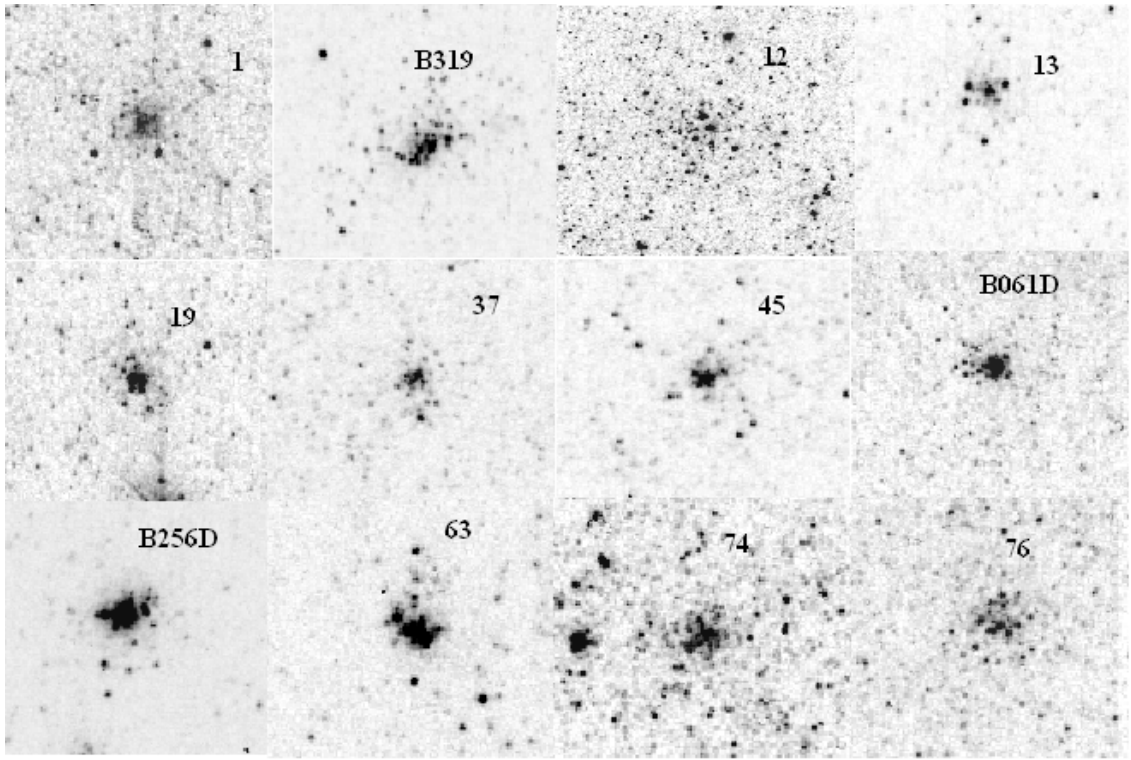

Figure 2. Images of 12 of the brightest clusters in the sample. Each small field is 7 arcsec on a side, except for cluster 12, for which the sides are 14 arcsec. The images are from the F450W filter and the WF camera.

### 3. DATA REDUCTION

**3.1 Integrated Photometry**

We determined integrated magnitudes and colors of the clusters using a photometric program written by Krienke in IDL and described in detail in KHI (2007). Magnitudes in the HST photometric system were calibrated according to the results of Holtzman et al. (1995). The program determines the cluster properties within a contour chosen to include most of the light, but omitting any bright foreground stars. The critical feature of the photometry is determining the background surface brightness (the "sky"). Because many of the clusters have both a low surface brightness and a significant size, the M31 background is often a significant fraction of the measured signal. Our program measures both a probable background level and determines the uncertainty of it by sampling several (10 to 24) similarly-dimensioned fields on the image These data are refined by Chauvenet criteria, rejecting samples with less than 0.02 probability of belonging to the set. The average of the remaining values of the background is then flux subtracted from the total flux within the cluster contour. The correction to the magnitudes due to the background subtraction was usually several tenths of a magnitude, but in some cases, where the cluster surface brightness was especially faint compared to the background, it reached values as large as 2 magnitudes (see Figure 3). Clearly, the background correction is an important element in this photometry and it is essential that it and its uncertainty be evaluated carefully. The photometric uncertainties provided in Figure 4 and Table 1 include that of the background, which in some cases dominates the uncertainty.

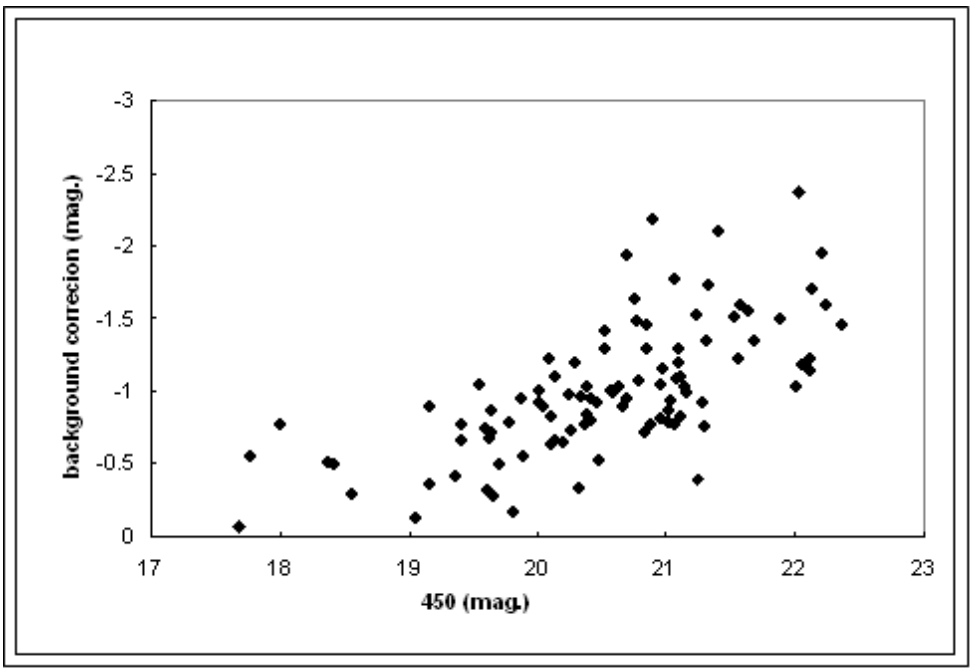

Figure 3. The background corrections plotted against the corrected integrated F450 magnitudes of the clusters. Magnitudes are not reddening-adjusted.

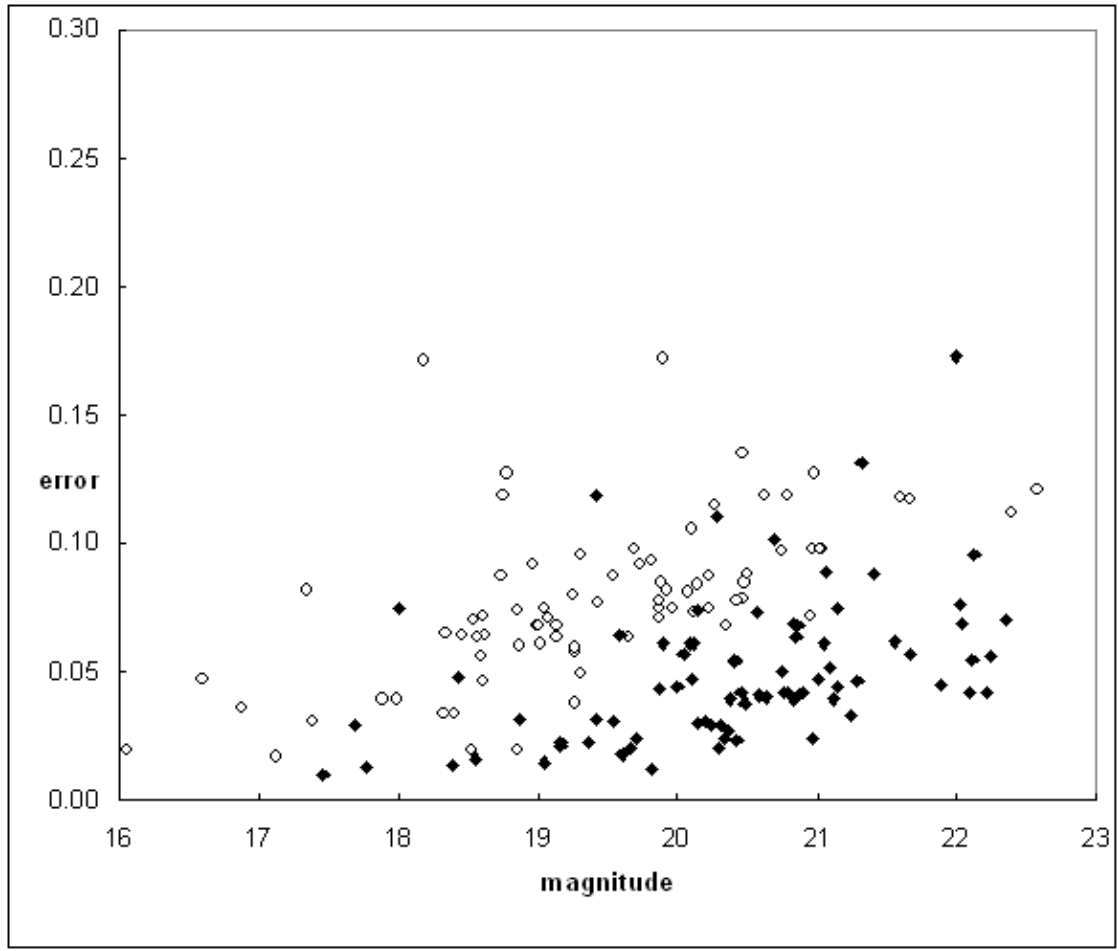

Figure 4. Photometric errors derived from the measurements of the integrated magnitudes, uncorrected for reddening. Filled symbols are for the F450 data and open symbols are for the F814 data.

**3.2 Stellar Photometry**

We carried out two independent programs of stellar photometry of the clusters. In one case, the entire WFPC2 images of each field were measured at Bologna as part of the luminous young clusters program. The details of that photometry are given in Paper I (Perina et al. 2009). For this paper we have extracted from the Bologna database the magnitudes and colors of stars within our outline of a cluster's boundary. Following the practice of Perina et al. (2009), we provide HST Vega magnitudes as measured in the two filters, which we refer to in the following as "F450" and "F814".

A second photometric program was carried out in Seattle using a program developed by one of us (OKK), based on DAOPHOT (Stetson 1994) and written within IDL. It was adjusted to allow us to measure stars in the more crowded central areas of clusters, where there often are bright stars, frequently including the brightest main sequence stars in the cluster. Without at least approximate photometry of these stars, we would be missing important information about the ages of the clusters. Zero points were adopted from Holtzman et al. (1995). PSFs were derived from several bright, well-separated stars in the field.

A comparison of the magnitudes and colors of the two sets of photometry showed good agreement. We identified stars in common by using both magnitudes and positions, finding that most bright stars were easily identified, while for faintest stars there was sometimes an ambiguity. For stars with F450 magnitudes brighter than 23.0 the mean differences (Bologna-Seattle) were -0.12 ± 0.05 magnitudes in F450 and -0.13 ± 0.11 magnitudes in F814. At fainter magnitudes, where the photometry is strongly affected by crowding and by the short exposures of the images, the dispersion is larger. We have adjusted the Seattle photometry to the Bologna system by using the above offsets

## 4. PROPERTIES OF THE CLUSTERS

### 4.1 The Cluster Catalog

Table 1 provides the positions, integrated magnitudes and integrated colors of the clusters. Five` of the clusters were found to have been identified previously according to the Revised Bologna Catalog of M31 Globular Clusters (Galleti et al. 2004, hereafter RBC). One of them, DAO84, was identified as a possible galaxy by Caldwell et al. (2009), but our images show a clearly-defined star cluster. Additionally, one coincides with an open cluster identified in KHI (2007) and one to a cluster

**Table 1**
Star Clusters of the Survey

| Name KHM31- | RA (J2000) | Dec. | F450 | err | F450 – F814 | err | |
|---|---|---|---|---|---|---|---|
| 22 | 9.99416 | 40.59044 | 20.36 | 0.03 | 1.38 | 0.07 | |
| 1 | 10.00226 | 40.59630 | 20.00 | 0.04 | 1.48 | 0.05 | |
| B319 | 10.01277 | 40.56638 | 17.77 | 0.01 | 0.89 | 0.04 | |
| WH | 10.03147 | 40.58568 | 20.75 | 0.05 | 0.64 | 0.09 | |
| 2 | 10.05996 | 40.47970 | 21.10 | 0.05 | 0.11 | 0.12 | *y |
| 3 | 10.06724 | 40.46574 | 20.87 | 0.07 | 0.72 | 0.11 | y |
| 4 | 10.07673 | 40.46278 | 20.23 | 0.03 | 0.93 | 0.06 | y |
| 5 | 10.08475 | 40.47733 | 21.29 | 0.05 | 0.81 | 0.10 | y |
| 6 | 10.09359 | 40.46366 | 22.10 | 0.04 | 0.50 | 0.13 | y |
| 7 | 10.10565 | 40.61191 | 21.23 | 0.13 | -1.01 | 0.18 | *y |
| 8 | 10.12093 | 40.60816 | 20.31 | 0.03 | 0.67 | 0.07 | y |
| 9 | 10.12172 | 40.62505 | 20.68 | 0.08 | 0.30 | 0.13 | *y |
| 10 | 10.12880 | 40.62470 | 20.26 | 0.04 | 0.01 | 0.11 | *y |
| 11 | 10.13828 | 40.61543 | 21.08 | 0.10 | 1.06 | 0.15 | |
| 12 | 10.14448 | 40.61308 | 18.00 | 0.08 | 1.42 | 0.09 | y |
| 13 | 10.15506 | 40.65390 | 19.36 | 0.02 | 1.47 | 0.05 | y |
| 14 | 10.15727 | 40.66958 | 20.83 | 0.04 | 1.71 | 0.06 | y |
| 15 | 10.17087 | 40.65345 | 20.96 | 0.06 | 0.74 | 0.11 | * |
| B014D | 10.25410 | 41.10937 | 19.60 | 0.02 | 1.63 | 0.04 | |
| 16 | 10.25739 | 41.12103 | 21.01 | 0.05 | 1.14 | 0.09 | y |
| 17 | 10.26360 | 41.11692 | 21.11 | 0.04 | 1.15 | 0.08 | |
| 18 | 10.27091 | 41.11649 | 20.42 | 0.02 | 1.16 | 0.04 | y |
| 19 | 10.27805 | 41.12904 | 19.41 | 0.12 | 1.23 | 0.21 | y |
| 20 | 10.31100 | 41.11747 | 22.03 | 0.08 | 1.23 | 0.14 | y |
| 21 | 10.32247 | 41.11345 | 20.69 | 0.10 | 1.95 | 0.16 | y |
| 22 | 10.32486 | 41.10686 | 21.40 | 0.09 | 1.18 | 0.12 | y |
| 23 | 10.32638 | 41.09547 | 21.88 | 0.05 | 1.60 | 0.10 | |
| 24 | 10.40369 | 40.72710 | 21.31 | 0.04 | -0.36 | 0.12 | *y |
| 25 | 10.40514 | 40.68031 | 20.56 | 0.07 | 1.42 | 0.10 | y |
| 26 | 10.41120 | 40.73322 | 18.55 | 0.02 | 0.21 | 0.07 | *y |
| 27 | 10.41445 | 40.67577 | 19.81 | 0.01 | 1.11 | 0.03 | y |
| 28 | 10.41904 | 40.72756 | 21.63 | 0.03 | -0.95 | 0.12 | *y |
| 29 | 10.42279 | 40.66916 | 20.19 | 0.03 | 0.92 | 0.07 | y |
| 30 | 10.42782 | 40.71453 | 19.66 | 0.02 | 0.69 | 0.07 | y |
| 31 | 10.43303 | 40.71460 | 21.08 | 0.04 | 0.12 | 0.11 | * |
| 32 | 10.43314 | 40.71762 | 21.09 | 0.05 | 1.22 | 0.09 | y |
| 33 | 10.43358 | 40.71122 | 20.89 | 0.04 | 2.04 | 0.09 | |
| 34 | 10.43870 | 40.72325 | 20.38 | 0.04 | 1.33 | 0.08 | y |
| 35 | 10.44996 | 40.71653 | 20.79 | 0.04 | 0.70 | 0.11 | y |
| 36 | 10.45031 | 40.69453 | 21.05 | 0.06 | 0.59 | 0.10 | y |
| 37 | 10.45168 | 40.69946 | 19.16 | 0.02 | 0.38 | 0.07 | y |
| 38 | 10.45521 | 40.72142 | 20.66 | 0.04 | 0.27 | 0.10 | *y |
| 39 | 10.45635 | 40.73367 | 21.08 | 0.26 | | | y |
| 40 | 10.46038 | 40.70244 | 20.58 | 0.04 | 0.69 | 0.09 | y |
| 41 | 10.51435 | 40.76969 | 20.14 | 0.03 | 1.53 | 0.08 | |
| 42 | 10.51689 | 40.74818 | 21.25 | 0.03 | 0.82 | 0.09 | y |
| 43 | 10.52399 | 40.77104 | 21.15 | 0.04 | 1.22 | 0.09 | y |
| 44 | 10.52901 | 40.76606 | 20.84 | 0.07 | 1.58 | 0.09 | y |
| 45 | 10.52987 | 40.76940 | 19.17 | 0.02 | 0.71 | 0.07 | y |
| 46 | 10.53052 | 40.77541 | 20.95 | 0.04 | 0.37 | 0.10 | *y |
| 47 | 10.53562 | 40.77516 | 19.70 | 0.02 | 0.71 | 0.07 | y |
| 48 | 10.55479 | 40.82819 | 20.63 | 0.04 | 1.20 | 0.09 | |
| 49 | 10.57024 | 40.81240 | 20.76 | 0.04 | 1.22 | 0.10 | y |
| 50 | 10.57764 | 40.81500 | 22.11 | 0.06 | 1.08 | 0.11 | |
| 51 | 10.57851 | 40.81922 | 19.89 | 0.06 | 1.35 | 0.09 | |
| B061D | 10.63578 | 41.36173 | 19.41 | 0.03 | 0.67 | 0.09 | * |
| 52 | 11.10224 | 41.25305 | 20.34 | 0.02 | 1.73 | 0.05 | |
| 53 | 11.11621 | 41.23792 | 20.96 | 0.02 | 2.11 | 0.03 | |
| 54 | 11.12238 | 41.23356 | 22.21 | 0.04 | 1.86 | 0.08 | |
| 55 | 11.22630 | 41.88489 | 21.28 | 0.04 | 0.19 | 0.10 | * |
| 56 | 11.23180 | 41.91120 | 20.29 | 0.02 | 0.92 | 0.04 | |
| 57 | 11.23438 | 41.89684 | 22.04 | 0.07 | 2.31 | 0.12 | |
| 58 | 11.23474 | 41.89572 | 20.11 | 0.06 | 1.15 | 0.11 | y |
| 59 | 11.23536 | 41.88171 | 20.45 | 0.04 | 2.06 | 0.05 | |
| 60 | 11.23619 | 41.91635 | 20.41 | 0.05 | 1.80 | 0.08 | |
| 61 | 11.24062 | 41.89716 | 22.12 | 0.10 | 1.38 | 0.14 | y |
| B256D | 11.24448 | 41.91018 | 17.57 | 0.02 | 1.58 | 0.03 | |
| 62 | 11.24560 | 41.89819 | 20.09 | 0.06 | 0.84 | 0.10 | y |
| 63 | 11.24637 | 41.91047 | 19.05 | 0.02 | 1.93 | 0.02 | |
| 64 | 11.24650 | 41.91050 | 18.87 | 0.03 | 1.88 | 0.05 | |
| 65 | 11.24744 | 41.89167 | 21.55 | 0.07 | -0.84 | 0.13 | *y |
| 66 | 11.24854 | 41.90391 | 20.21 | 0.09 | 1.43 | 0.12 | y |
| 67 | 11.24969 | 41.93580 | 20.85 | 0.06 | 0.78 | 0.10 | y |
| 68 | 11.24973 | 41.90117 | 21.32 | 0.13 | 1.06 | 0.17 | y |
| 69 | 11.25109 | 41.90682 | 21.06 | 0.09 | 1.17 | 0.19 | y |
| 70 | 11.25216 | 41.88646 | 20.48 | 0.04 | 1.17 | 0.10 | y |
| 71 | 11.25366 | 41.88541 | 19.87 | 0.04 | 0.85 | 0.08 | y |
| 72 | 11.25606 | 41.89460 | 21.76 | 0.13 | 0.76 | 0.17 | |
| 73 | 11.25914 | 41.91537 | 19.97 | 0.04 | 1.31 | 0.07 | |
| 74 | 11.26204 | 41.89759 | 20.52 | 0.09 | 0.67 | 0.12 | y* |
| 75 | 11.26219 | 41.90101 | 20.38 | 0.08 | -0.08 | 0.16 | *y |
| 76 | 11.26942 | 41.89441 | 20.02 | 0.06 | 1.03 | 0.11 | |
| 77 | 11.28053 | 41.90742 | 21.67 | 0.06 | 0.83 | 0.11 | y |
| 78 | 11.28957 | 41.91235 | 21.56 | 0.06 | 0.61 | 0.10 | y |
| 79 | 11.29089 | 41.91942 | 20.10 | 0.05 | 1.50 | 0.07 | y |
| 80 | 11.43302 | 41.72510 | 19.63 | 0.03 | 0.55 | 0.08 | y |
| 81 | 11.45692 | 41.71174 | 22.35 | 0.07 | 1.85 | 0.11 | |
| 82 | 11.45853 | 41.70832 | 22.23 | 0.06 | 1.61 | 0.13 | |
| DA084 | 11.46799 | 41.71365 | 19.59 | 0.06 | 0.81 | 0.14 | |

NOTES: Objects with asterisks have uncertain colors because of a low ratio of signal to galaxy background in the F814W image. Objects with "y" have CMDs indicating young ages, less than ~$5 \times 10^8$ years.

discovered by Williams and Hodge (2001b). Only two of the previously-identified clusters, B319 and KH22, had published magnitudes in B and only B319 had previously-published magnitudes in both B and I. We transformed our magnitudes to Johnson-Cousins B and I for comparison. The average difference (previous – this paper) in B was found to be 0.16 mag. and the difference in I is 0.18 mag.

As a ground-based check on the HST photometry, one of us (JS) determined the integrated magnitudes and colors of 16 of the brighter clusters from the SDSS database. Measures were obtained in the SDSS system (u,g,r,i,z) and transferred to B and I in the J-C system. All measures used a circular aperture with a radius of 4 arcsec. The measures produced data that agreed fairly well, with mean differences (CfA-Seattle) of $\Delta B$ = -0.24 +/- 0.39 and $\Delta(B - I) = 0.23$ +/- 0.14. Experiments with HST photometry using a 4 arcsec aperture indicated that the differences are probably caused at least partly by nearby bright stars that were avoided by the original HST photometry, which used smaller apertures.

## 4.2 The Integrated Cluster Color-Magnitude Diagram

Figure 5 shows the color-magnitude diagram (hereafter CMD) of the present sample (we include in this diagram and in Figure 6 two clusters from the main target program, which were found serendipitously on the WF frames). It closely resembles the two diagrams published for similar samples of M31 clusters by KH I and II (2007, 2008), though with different filter pairs. The mean absolute magnitude for the cluster sample plotted is $M(F450)_0 = -4.59$ and the mean unreddened color is $(F450-F814)_0 = 0.67$.

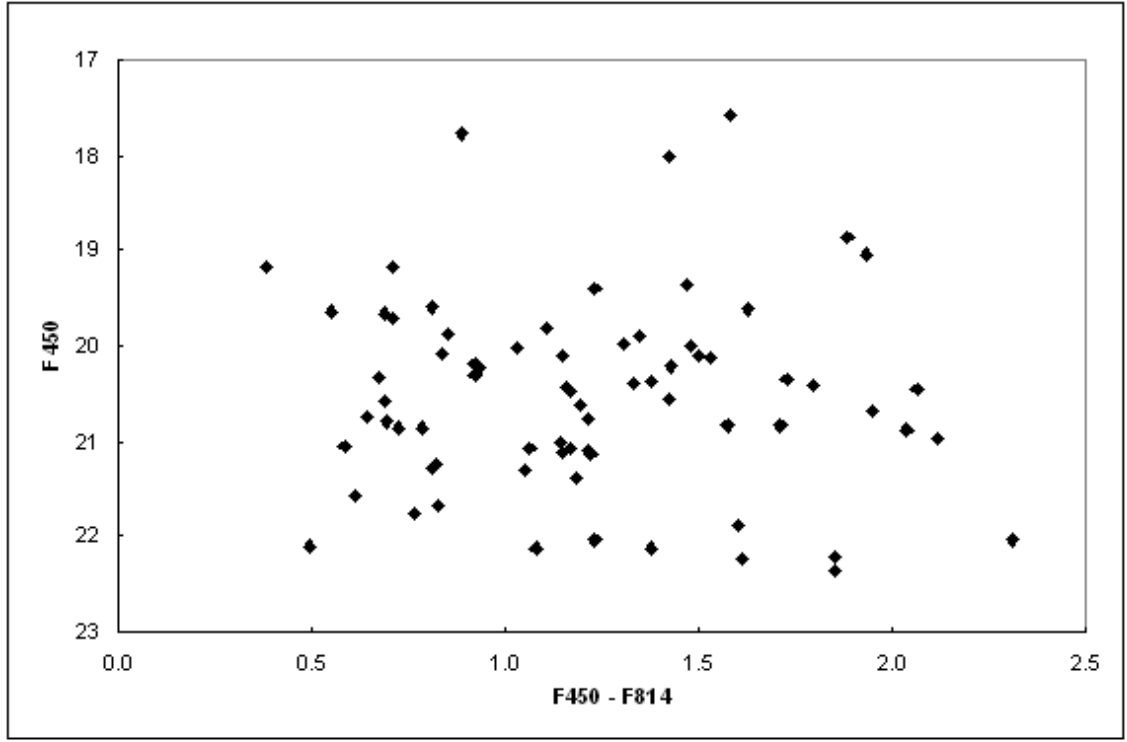

Figure 5. The color-magnitude diagram for the integrated colors and magnitudes of clusters in this survey. The plot shows observed values, before corrections for reddening.

The clusters are nearly uniformly distributed over the diagram, but with a mild concentration at about $F450 = 21$ and $F450 - F814 = 1$. For reference, a cluster with observed values of $F450 = 21.0$ and $F450 – F814 = 1.0$ will have an age of about ~70 Myr and a mass of 450 solar masses, assuming a Salpeter stellar luminosity function and Giardi (2006) population models. But note that the age-color diagram is multi-valued at these colors (see Section 5.2).

The mean size of the isophotal radii of all clusters was 1.61 arcsec (6.12 pc).

**4.3 The Integrated Cluster Luminosity Function**

The luminosity function of the clusters is shown in Figure 6, where the magnitudes are corrected for extinction, assuming a mean reddening of $F450 – F814$ of 0.51 (see Section 6). The shape of the luminosity function is approximately Gaussian, with a maximum at

M(F450)(0) = -4.2. All three samples show an enhanced frequency at the bright end, compared to a symmetrical curve. Artificial cluster tests on the WFPC2 HST images in KHI indicated that much of the turn-down at faint magnitudes results from detection limits. It is not yet clear what the shape of the true luminosity function is at such faint limits. While KHI suggested that the luminosity function may continue to rise, at least to M(F450) = -1, similar HST searches for faint clusters in the SMC have produced contrary results (Rafelski and Zaritsky, 2005). In any case, the luminosity function at the faint end is a complicated product of selection effects, evolutionary fading rates and dynamical disruption (Hunter et al. 2003).

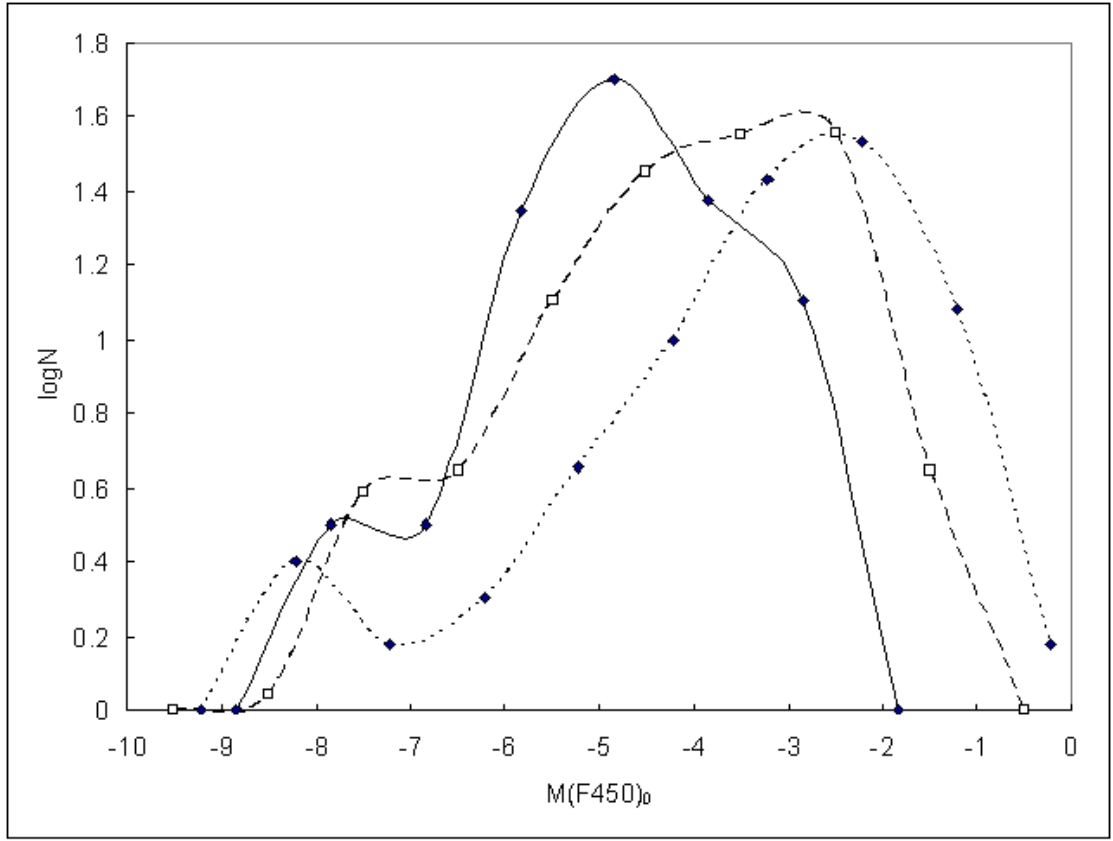

Figure 6. The luminosity function for the clusters of this survey (solid line) compared to that of KHI (2007) (dotted line) and KHII (2008)(dashed line). The latter two are normalized to the total number of clusters in the present survey.

**4.4 Individual Cluster CMDs**

As described in Section 3.2, we measured stellar CMDs for all clusters. Most diagrams looked reasonable, but not all of the clusters were well-enough resolved to allow meaningful interpretation. Especially for the faintest clusters, the number of stars on the F814 frame was often quite small, on the order of 5 to 10.

Figure 7 shows the CMDs for 10 clusters for which the CMDs show a well-defined main sequence. These clusters show a main sequence with F450 – F814 near 0.5 and with the tip of the main sequence in the range with F450 magnitudes = 20 to 24. The CMDs in Figure 7 have been adjusted for reddening (see Section 5.1).

Table 2 lists the clusters for which it was possible to determine age and reddening by comparison with the Girardi models. The quoted uncertainties indicate the extreme limits of acceptable fits judged by eye.

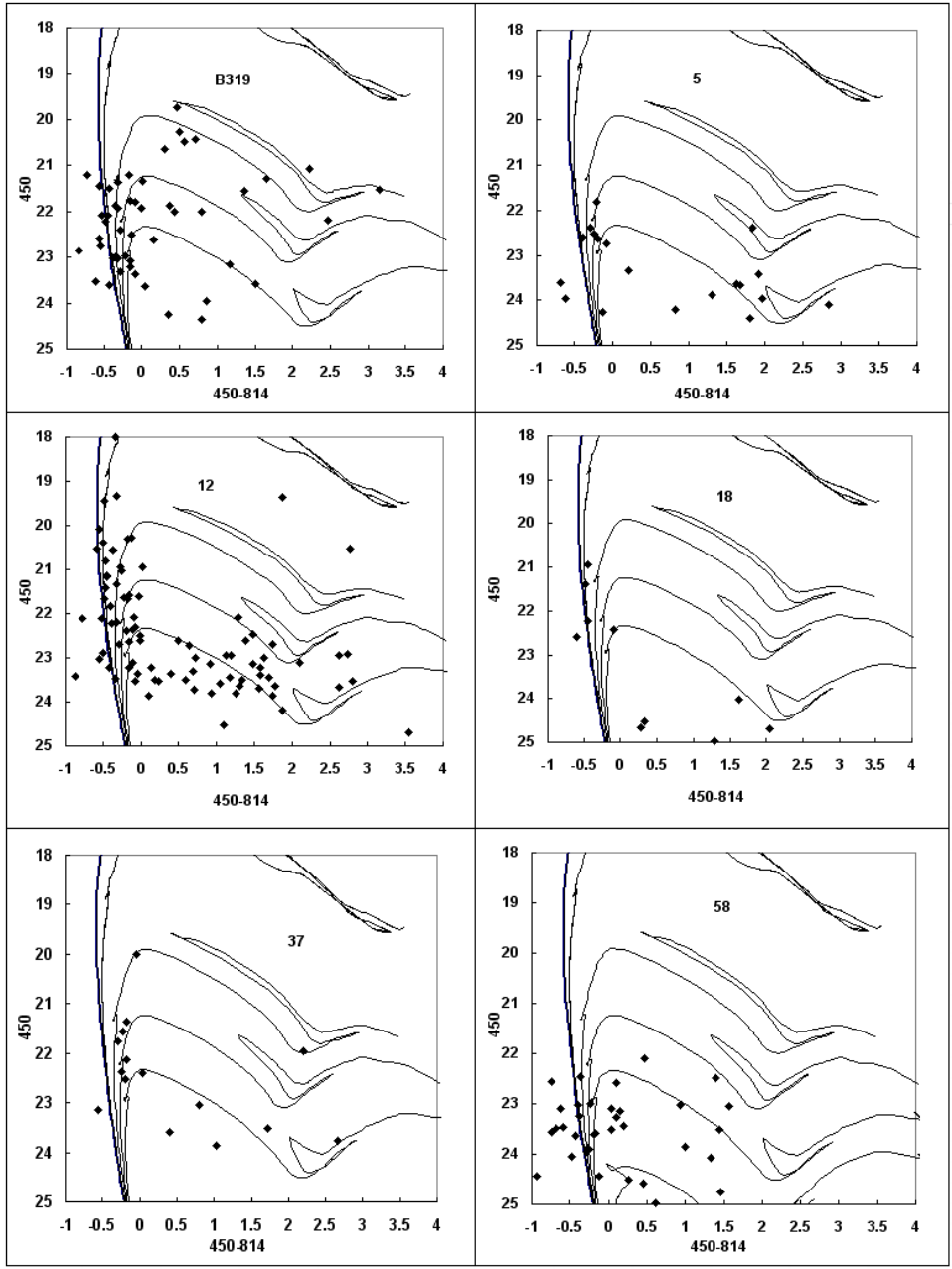

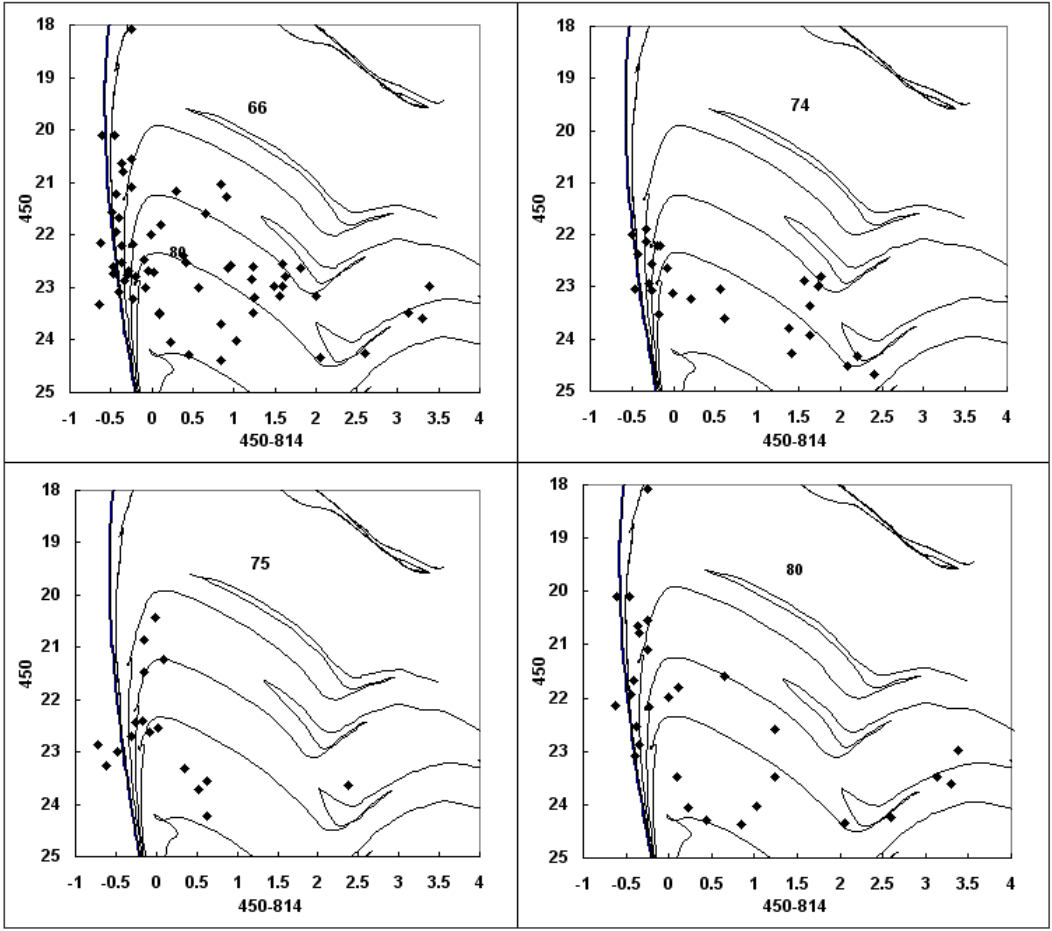

Figure 7. Color-magnitude diagrams for 10 young clusters with well-defined main sequences, fitted by eye to Girardi (2006) isochrones for solar abundance and ages with log(age) of 7.0, 7.6, 8.0, 8.25, and 8.7 years.

One of the clusters, B319 (also known as G44) has been studied previously using other HST images (Williams and Hodge, 2001a). The present CMD is shallower and it covers only the central region of B319, but the two CMDs are morphologically similar. We cannot usefully make detailed comparisons because the Williams and Hodge data were taken with different filters (F 336W, F439W and F555W).

A careful inspection of the CMDs of the clusters and their surrounding fields shows that the degree of contamination of the cluster MS by field stars is negligibly low and does not affect our estimates of age and reddening.

**Table 2**
Characteristics of Cluster CMDs with Well-Defined Main Sequences

| cluster no. | log age(yrs) | uncertainty | E(F450-F814) | uncertainty |
|---|---|---|---|---|

| Cluster | | | | |
|---|---|---|---|---|
| KH22 | 7.6 | 0.35 | 0.4 | 0.15 |
| B319 | 7.6 | 0.5 | 0.5 | 0.25 |
| 3 | 7.5 | 0.45 | 0.8 | 0.2 |
| 5 | 8.0 | 0.6 | 0.5 | 0.3 |
| 8 | 7.5 | 0.35 | 0.55 | 0.2 |
| 11 | 7.3 | 0.6 | 0.5 | 0.2 |
| 12 | 7.6 | 0.6 | 0.55 | 0.25 |
| 13 | 7.1 | 0.5 | 0.85 | 0.8 |
| 18 | 7.1 | 0.35 | 0.5 | 0.2 |
| 34 | 8 | 0.45 | 0.65 | 0.25 |
| 37 | 7.9 | 0.35 | 0.5 | 0.25 |
| 45 | 7.8 | 0.3 | 0.5 | 0.15 |
| B061D | 7.8 | 0.6 | 0.5 | 0.15 |
| 58 | 7.6 | 0.2 | 0.8 | 0.15 |
| 62 | 8.0 | 0.2 | 0.25 | 0.15 |
| 68 | 7.8 | 0.3 | 0.82 | 0.15 |
| 74 | 8.1 | 0.3 | 0.65 | 0.15 |
| 75 | 7.8 | 0.5 | 0.5 | 0.25 |
| 80 | 7.1 | 0.45 | 0.75 | 0.15 |

## 5. AGES AND REDDENINGS

### 5.1. From the CMDs

For clusters with sufficiently well defined sequences of stars, especially young clusters with narrow main sequences, it was possible to determine approximate reddenings and ages. Based on the case for vdB0 (Perina et al. 2009), we assumed that these young clusters are characterized by solar abundances. We compared the observations with evolutionary model isochrones made available from the Padua webpage (Girardi 2006) and determined the offset by eye, providing approximate values of age and reddening (Table 3). Because of the faintness of the magnitudes, the crowding and the sparseness of the CMDs, these values have fairly large uncertainties, as quoted in the table. Within the accuracy of the fitting and if our assumption of solar abundances is correct, the fits provide individual reddenings for the selected clusters, which range from E(F450-F814) = 0.25 to 0.85, with a mean uncertainty of 0.23. The average reddening for this sample is 0.59 with a standard deviation of 0.21 magnitudes. Selection effects, of course, severely limit our sample of clusters with bright main sequences to the youngest clusters in the sample; most are younger than 200 million years.

For the remaining clusters in the sample, the color-magnitude diagrams are difficult to interpret in terms of ages and reddenings except in approximate terms. Table 1 notes those clusters that have significant numbers of stars in the blue section of their CMDs to indicate that they are younger than a few times $10^8$ years. Most of the remaining clusters are older, as is also indicated by their integrated colors.

### 5.2 From the Integrated Cluster Photometry

Integrated colors of open clusters can be used to estimate cluster ages by comparison with theoretical models. There are a number of problems with this procedure in our case:

a) the colors are intrinsically uncertain because of the spatially-variable brightness and color of the M31 background, which is the major source of the photometric uncertainty.
b) the theoretical models show a dependence on the elemental abundances, which are unknown.
c) for young small-mass clusters, the colors depend on small number statistics in the presence or absence of the most luminous blue stars or a few red giants (see Frogel, Cohen and Persson, 1983 and Cervino and Luridiana, 2004, for quantitative treatments of this problem).
d) different theoretical models, even for the same abundances, give different relationships for the age-color diagram.
e) for the colors used in this program (F450 and F814), the change with color for young clusters ($<2\times10^8$ yrs.) is multi-valued for some regimes and is generally smaller than the measurement uncertainties (Figure 8).

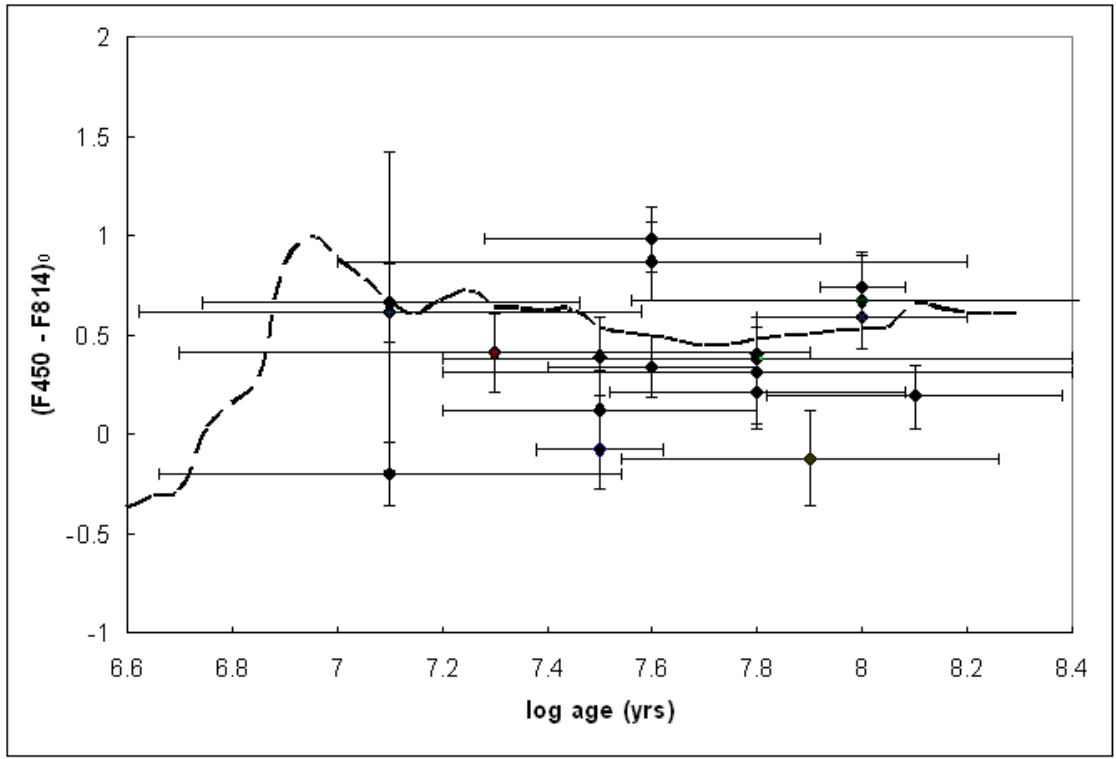

Figure 8. Ages and reddening-corrected colors determined from MS fitting compared to the theoretical age-color relationship for young clusters (Girardi 2006).

In spite of these difficulties, it is possible to estimate approximate ages from the colors and, for the younger clusters, the average reddening. Figure 8 shows the colors of the clusters with well-defined main sequences compared to the theoretical colors for single-age populations with solar abundances (Girardi 2006). The colors plotted are the measured colors corrected for reddening and the reddening and ages are those determined from main sequence fitting. The colors cluster close to the theoretical distribution but are clearly offset to the blue. This may be due to abundances that are different from our assumption of solar abundance ratios. Alternatively, if we assume the offset to be due to overestimation of reddening, then the best fit to the models is for a mean reddening 0.085 magnitudes smaller than derived from the MS fitting and gives a mean reddening of E(F450-F814) = 0.50 (this corresponds to E(B – V) = ~0.25). For our complete sample we adopt this value for the mean reddening.

For ages of clusters older than ~300 million years the theoretical curve is single-valued and fairly sensitive to the measured colors. Because of our shallow exposures, it is not possible to derive ages from CMDs for these clusters, but we can estimate ages from colors, if we assume a mean reddening and a particular model set and abundance. Table 3 provides approximate ages for the clusters with colors redder than (F450 – F814) = 1.0. These data are calculated with a mean reddening of E(F450-F814) = 0.50 and use the models provided by Girardi (2006). Formal errors of the colors correspond to approximately an uncertainty of 0.10 in log age, but the true uncertainties of the ages are considered to be much larger, for the reasons outlined at the beginning of this section. The reddest clusters in the sample have reddening-corrected colors of F450-F814 = ~1.8, which corresponds to an age of approximately $1.5 \times 10^9$ years.

## Table 3
### Ages for Older Clusters Based on Integrated Colors

| Name | logage,yr |
|---|---|
| 1 | 8.63 |
| 14 | 8.77 |
| B014D | 8.72 |
| 16 | 8.29 |
| 17 | 8.30 |
| 19 | 8.38 |
| 20 | 8.38 |
| 21 | 8.94 |
| 22 | 8.33 |
| 23 | 8.70 |
| 25 | 8.56 |
| 27 | 8.25 |
| 32 | 8.37 |
| 33 | 8.97 |
| 34 | 8.50 |
| 41 | 8.64 |
| 43 | 8.37 |
| 44 | 8.68 |
| 48 | 8.37 |
| 49 | 8.38 |
| 50 | 8.22 |
| 51 | 8.50 |
| 52 | 8.79 |
| 53 | 9.04 |
| 54 | 8.87 |
| 57 | 9.22 |
| 59 | 8.99 |
| 60 | 8.84 |
| 61 | 8.53 |
| B256D | 8.68 |
| 63 | 8.92 |
| 64 | 8.88 |
| 66 | 8.57 |
| 69 | 8.32 |
| 70 | 8.32 |
| 73 | 8.46 |
| 76 | 8.15 |
| 79 | 8.63 |
| 81 | 8.87 |
| 82 | 8.71 |

## 5.3 The Age Distribution.

We have suggested above that the color-magnitude diagram of integrated magnitudes (Figure 5) indicates that the clusters are not distributed uniformly in age. To examine the age distribution we have combined the age data for the young clusters based on main sequence fitting with that for older clusters based on colors. Figure 9 shows the distribution for our sample of 82 clusters. The number falls off rapidly with age, approximately exponentially. A least squares linear fit gives

$$\log(N) = -1.625\log(t) + 11.676$$

Also shown in Figure 9 is a similar curve for the clusters in KHI, where the number has been normalized to adjust for that survey's larger sampling area. The two agree within their errors, though there is a suggestion of a small difference in slope, which is possibly caused by the shallower exposure times of the present survey, which probably missed a larger fraction of older clusters.

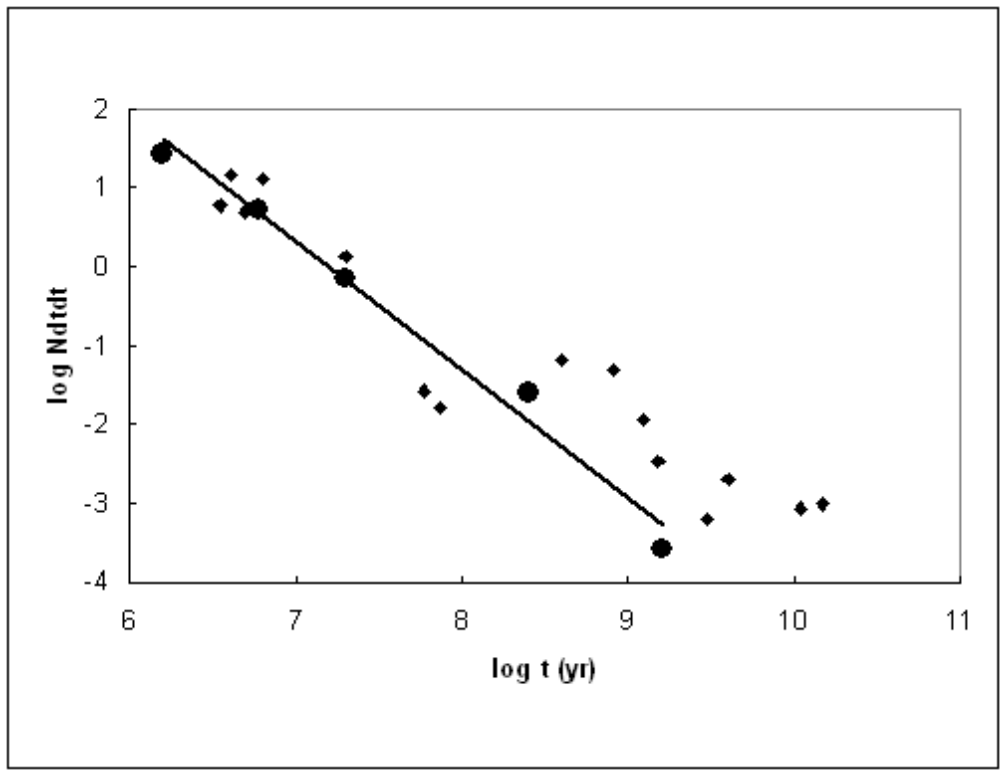

Figure 9. The age distribution for the clusters in this survey (large circles) compared to that reported in KHI (diamonds). The line is a least squares linear fit to this paper's data.

As discussed briefly in KHI and in a large and diverse recent literature, these kinds of data are useful for determining the survival rate of clusters in a galaxy's gravitational field (e.g., Kruijssen and Lamers, 2008, Gieles, et al. 2006, Chandar, Fall and Whitmore 2006, Lamers and Gieles, 2006 and many others). Before such use can be made of the data, however, it is necessary to know both the rate of evolutionary fading of the clusters and the detection efficiency of the survey. We note that the fading rate is dependent on the abundances, which are unknown, and the detection efficiency is dependent on the exposure times, on the structural properties of the clusters and on the background surface brightness and its variability. To determine the detection efficiency for a collection of such faint and varied clusters would require a much larger sample, as each of the determining factors would need to be explored. In view of these difficulties, we believe that the current survey is not appropriate for deriving a tidal destruction rate for M31 clusters.

## 6. SUMMARY

This paper supplements the HST/WFPC2 Survey of Luminous Young Clusters in M31, which examines the nature of 19 globular-like objects that are anomalously blue. Our search for other, less luminous clusters on the images has produced a catalog of 89 clusters, 82 of which are newly-identified.

We have obtained integrated magnitudes and colors of the clusters and have measured color-magnitude diagrams for their resolved stars. The absolute magnitudes of the clusters range from M(F450) = -8 to -2.5 and their colors indicate a large range of ages, from a few million to a few times $10^9$ years. The richest young clusters have well-defined main sequences that have been fitted to theoretical isochrones, providing ages ranging from approximately 12 million to 100 million years. The CMDs of these clusters indicate reddenings averaging E(F450-F814) = 0.59, with a dispersion of 0.21 magnitudes, while a comparison of integrated colors of a larger sample of the young clusters with theoretical population models indicates a somewhat smaller average reddening of 0.50 magnitudes. We derive a cluster luminosity function that shows a peak value of $M(F450)_0$ of -4.2 and which extends from values of -9 to -2. The least luminous clusters are among the faintest measured for clusters in LG galaxies. There is a suggestion of a small number of anomalously-luminous clusters at the bright end of the luminosity function. The distribution of the number of detected clusters with age shows a very steep gradient.

This paper was based on observations made with the NASA/ESA Hubble Space Telescope, obtained at the Space Telescope Institute, which is operated by the Association of Universities for Research in Astronomy, Inc. under NASA contract NAS 5-26555. These observations are associated with program GOI-10818 (P. I.: J. G. Cohen) and were partially funded under that program.